\documentclass[12pt]{article}

\usepackage{amsmath}
\usepackage{threeparttable}
\usepackage{braket}
\usepackage{graphicx}
\usepackage{dcolumn}
\usepackage{bm}
\usepackage{hyperref}
\usepackage{cleveref}
\usepackage[biblabel]{cite}
\usepackage{authblk}
\usepackage{setspace}
\usepackage[margin=1.2in,
]{geometry}
\title{On the inversion of isobaric-analogue states in mirror nuclei}
\author[1]{J.~Henderson\thanks{henderson64@llnl.gov}}
\author[2]{S.~R.~Stroberg\thanks{stroberg@uw.edu}}
\affil[1]{Lawrence Livermore National Laboratory, Livermore CA, USA}
\affil[2]{Department of Physics, University of Washington, Seattle WA, USA}

\begin{document}

\date{}

\maketitle



Isospin is an approximate symmetry in atomic nuclei, arising from the rather similar properties of protons and neutrons.
Perhaps the clearest manifestation of isospin within nuclei is in the near-identical structure of excited states in mirror nuclei: nuclei with inverted numbers of protons and neutrons~\cite{ref:Bentley_07}. 
Isospin symmetry, and therefore mirror-symmetry, is broken by electromagnetic interactions and the difference in the masses of the up and down quarks.
A recent study by Hoff and collaborators~\cite{ref:Hoff_20} presented evidence that the ground-state spin of \textsuperscript{73}Sr is different from that of its mirror, \textsuperscript{73}Br, due to an inversion of the ground- and first-excited states, separated by only 27~keV in the \textsuperscript{73}Br system.
In this brief note, we place this inversion within the necessary context of the past half-century of experimental and theoretical work, and show that it is entirely consistent with normal
behaviour, and affords no new insight into isospin-symmetry breaking. 
The essential point is that isospin-breaking effects due to the Coulomb interaction frequently vary from level to level within a given medium-mass nucleus by as much as 200 keV. Any level splitting smaller than this is liable to manifest a level inversion in the mirror partner which, absent disagreement with an appropriate nuclear model, does not challenge our understanding.
While we note the novelty of an inversion in nuclear ground states, we emphasize that in the context of isospin there is nothing specifically illuminating about the ground state, or a level inversion.

A great deal of experimental work has been performed in mid-mass nuclei~\cite{ref:ENSDF,ref:Spieker_19,ref:Ong_17,ref:Milne_16,ref:Davies_13}, in an effort to understand isospin-symmetry breaking through energy shifts in mirror nuclei.
We employ the data generated from that work to understand the magnitude of normal isospin-breaking effects.
(By ``normal'' we mean consistent with the scale of the leading isospin-breaking terms in the Hamiltonian.)
We define $E_{12}$ as the energy difference between two states in a given nucleus. The relevant quantity for state inversion in the mirror nucleus is the difference 
\begin{equation}
    \delta E_{12} = E_{12}^{N>Z} - E_{12}^{N<Z},
\end{equation}
where N and Z are the neutron and proton numbers, respectively. If $\delta E_{12}$ is negative and larger in magnitude than the level spacing in the neutron-rich partner, $E_{12}^{N>Z}$, an inversion will occur. The question is how common such an inversion might be.

\begin{figure}
\centerline{\includegraphics[width=\linewidth]{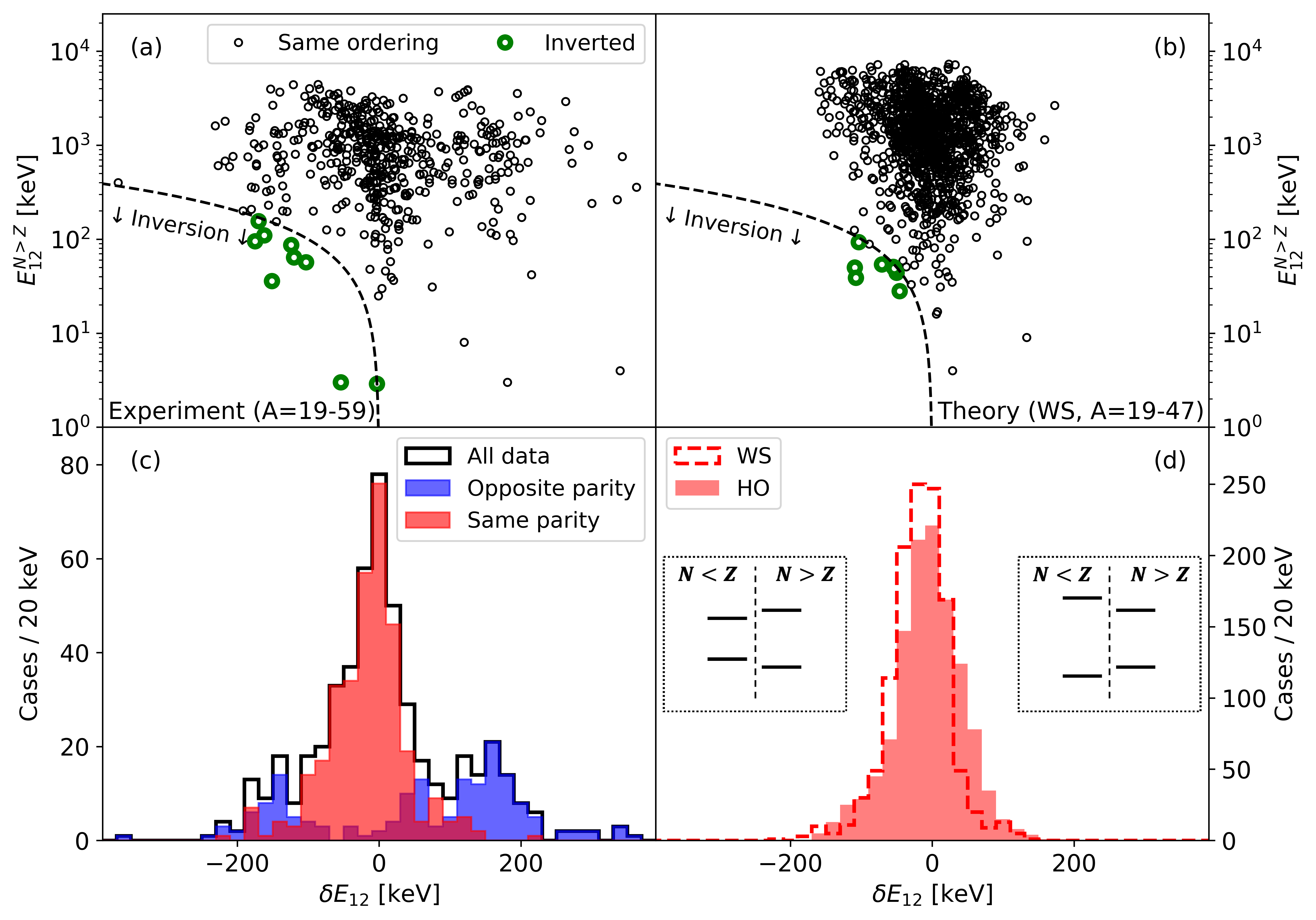}}
\caption{(Top) Experimental (a) and Woods-Saxon theoretical (b) energy splitting ($E_{12}^{N>Z}$) of the states in the $N>Z$ nucleus plotted against $\delta E_{12}$. Dotted lines indicate $E_{12}^{N>Z} = -\delta E_{12}$. Points below the lines result in a state inversion in translating to the $N<Z$ nucleus. (Bottom) Experimental (c) and theoretical (d) projections of (a) and (b), respectively. The experimental data are separated into like- and opposite-parity pairs. The theoretical calculations can only produce like-parity pairs. As shown schematically in the insets to (d), the effect of a negative $\delta E_{12}$ is to compress and potentially invert the level spacing in the $N$$<$$Z$ nucleus, while a positive value causes their spacing to expand.
}
\label{fig:dMED}
\end{figure}

Taking the available experimental data for odd-odd and odd-mass mirror pairs between $A$=19 and $A$=59, panel (a) of Fig.~\ref{fig:dMED} shows the energy splitting between the states in the $N$$>$$Z$ nucleus, $E_{12}^{N>Z}$ plotted against $\delta E_{12}$. States at or beyond the proton-separation energy are excluded from the analysis to avoid significant contributions from weak binding.
The dashed line shows $E_{12}^{N>Z} = -\delta E_{12}$, indicating the $\delta E_{12}$ required to overcome the energy splitting between the states. Twenty-four of the 508 state combinations satisfy $|\delta E_{12}|>E_{12}^{N>Z}$, of which nine undergo an inversion in transforming between the $N$$>$$Z$ and $N$$<$$Z$ systems.  

The histogram in panel (c) of Fig.~\ref{fig:dMED} shows the same $\delta E_{12}$ values, demonstrating that positive and negative values are approximately equally likely, with a characteristic scale of 100-200~keV.
Notably, there is a concentration of cases around $\delta E_{12} = 0$, indicating a similar energy difference between different IAS in the mirror pair. One possible explanation for this effect is that the two IAS have a similar underlying structure (e.g. they are members of the same rotational band), and thus experience similar changes in the transformation between mirror nuclei. In fact, we may expect that the data is biased towards states within the same rotational band, as such states are more likely to have a firm spin assignment. This would not impact our conclusions. 
Qualitatively, we assume the states of different parity within a given mirror pair will have somewhat different structure, and would therefore behave differently when transforming between mirror nuclei.
The experimental data fit well with this description: IAS-pairs with the same parity tend to have a significantly smaller $\left|\delta E_{12}\right|$, while pairs with opposite parity typically experience larger splittings of 100-200~keV.

\begin{figure}
\vspace{+5pt}
\centerline{\includegraphics[width=\linewidth]{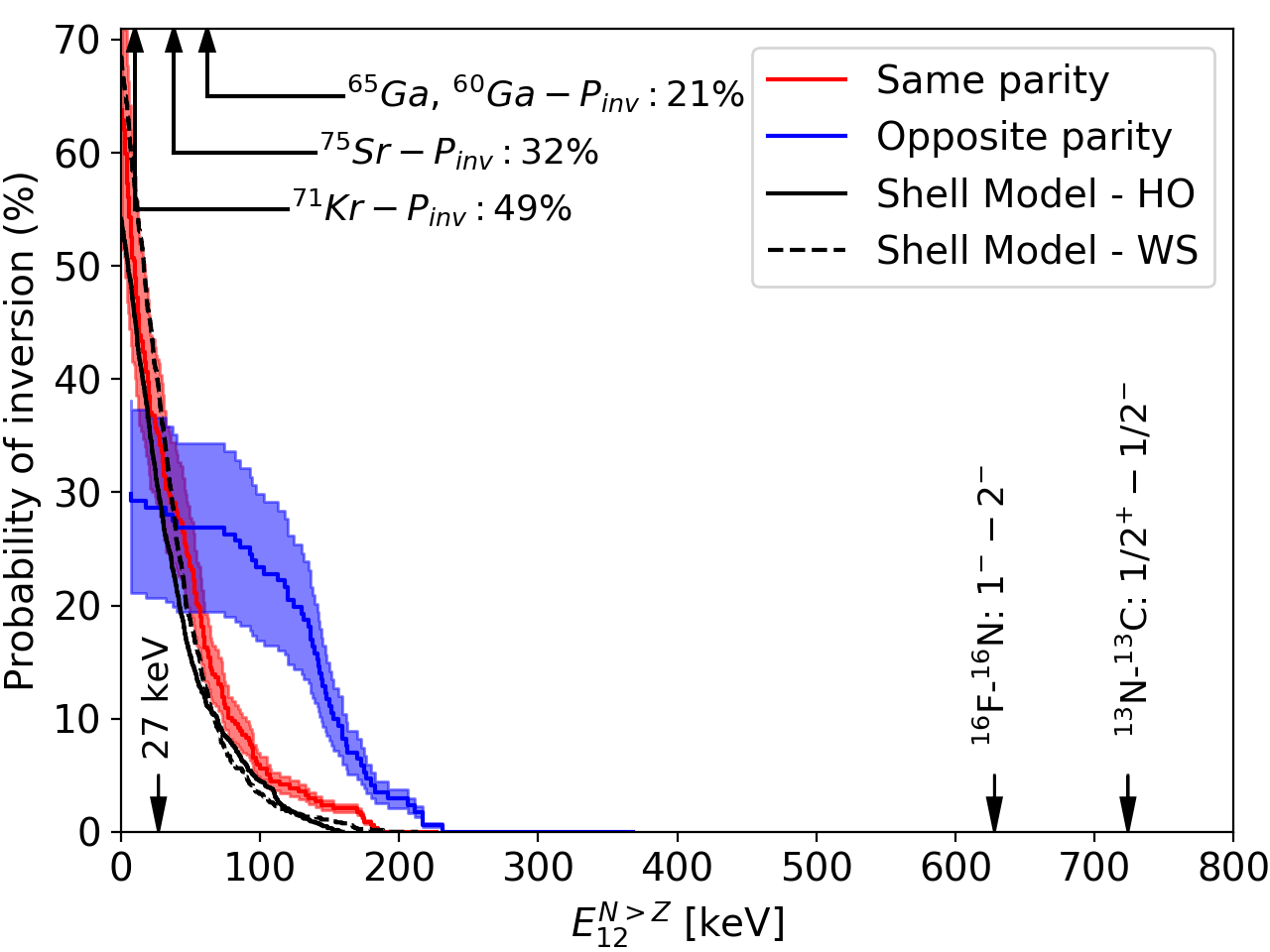}}
\caption{Calculated probability that a pair of states with energy difference $E_{12}^{N>Z}$ will invert.
Probabilities are shown for like- and opposite-parity pairs, as well as for the shell-model results. Assuming strong correlations between states within a nucleus, relative uncertainties are calculated as $\sqrt{N_p}/N_p$, where $N_p$ is the number of mirror-systems that exhibit same- and opposite-parity pairs. Also shown are two examples of observed splittings due to isospin-symmetry breaking phenomena: the $2^-$ and $1^-$ states in \textsuperscript{16}F-\textsuperscript{16}N, and the $1/2^{+}$ and $1/2^{-}$ states in the \textsuperscript{13}N-\textsuperscript{13}C mirror pair\cite{ref:Thomas_52,ref:Ehrman_51}. The 27 keV energy splitting in \textsuperscript{73}Br is highlighted. We also indicate other likely cases for ground-state inversion, but emphasize that these inversions would not provide more insight than any other mirror energy shifts with regards to isospin-symmetry breaking.}
\label{fig:Prob}
\end{figure}

Thus far we have relied entirely on experimental data, and one could imagine some exotic symmetry-breaking effects lurking in that same data.
Panels (b) and (d) of Fig.~\ref{fig:dMED} show the result of shell model calculations of $E_{12}^{N>Z}$ and $\delta E_{12}$ for $|N-Z|=1,3,5$ 
mirror pairs in the $sd$ and $f_{7/2}$ shells.
We employ the code NuShellX~\cite{ref:NushellX} and the phenomenological USDB~\cite{ref:USDB} and GXPF1A~\cite{ref:Honma_05} interactions, which conserve isospin. To compute $\delta E_{12}$, we evaluate the Coulomb potential $\frac{e^2}{r}$ in first order perturbation theory.
In general, the single-particle wave functions of phenomenological shell model interactions are not specified, and so we explore two choices.
In the first, labeled HO, we take harmonic oscillator wave functions with frequency $\hbar\omega = 45A^{-1/3}-25A^{-2/3}$~MeV\cite{ref:Blomqvist_68}.
In the second, labeled WS, we use eigenstates of a Woods-Saxon potential~\cite{ref:Suhonen_07} $V(r)= \left(V_0+V_1\frac{N-Z}{A}\right) f(r) + V_{\ell s}\frac{r_0}{\hbar^2 A^{1/3}}\langle \ell\cdot s\rangle \frac{d}{dr}f(r) + V_{\rm C}(r)$.
We diagonalize the one-body Hamiltonian in a basis spanned by 21 major harmonic oscillator shells, employing the same oscillator frequency as in the HO case.
The function $f(r)=\left[1+\exp(\frac{r-R_0}{a})\right]^{-1}$ is the standard Woods-Saxon form, and $V_{\rm C}(r)$ is the Coulomb potential due to a uniformly charged sphere of radius $R_0$.
We take $V_0=-50$~MeV, $V_1=-33$~MeV, $V_{\ell s}=-20$ MeV, $r_0=1.25$~fm, $a_0=0.65$~fm, $R_0=r_0 A^{1/3}$. As with the experimental data, the theoretical data were required to have excitation energies below the proton-separation energy.
It is clear that the Coulomb interaction is sufficient to produce the distribution of $\delta E_{12}$ values.
From a point-by-point comparison the WS choice has a root-mean-squared (rms) deviation from experiment of 53~keV, as compared to 88~keV for the HO choice, while the rms deviation between the two methods is 74~keV. This highlights the difficulty of extracting information about isospin-symmetry breaking forces from individual level shifts (especially small ones) due to ambiguities in the single-particle behaviour.
On the other hand, both methods produce very similar $\delta E_{12}$ distributions.

In the A=73 pair~\cite{ref:Hoff_20}, the ground- and first-excited states are described as belonging to different, near-degenerate intrinsic shapes.
It is not unreasonable to expect that the behaviour of these states will more closely resemble that of IAS-pairs with differing parity in Fig.~\ref{fig:dMED} than those of like-parity. To estimate the likelihood of a pair of IAS undergoing an energy inversion, Fig.~\ref{fig:Prob} shows the probability that a pair of states will invert based on the data in Fig.~\ref{fig:dMED}. Approximately 35~\% of like-parity IAS-pairs separated by 27 keV would be expected to invert. The same probability distribution is calculated based on the shell-model calculations and reproduces the like-parity distribution well. 

Figure~\ref{fig:Prob} also shows two famous cases of additional isospin-symmetry breaking phenomena, both arising in unbound states~\cite{ref:Thomas_52,ref:Ehrman_51,ref:Stefan_14}, with one resulting in the only other known case in which the ground states of a mirror pair differ.
Both of these cases significantly exceed the distribution of data shown in Fig.~\ref{fig:dMED}, demonstrating that there is a need to invoke additional isospin-symmetry-breaking effects (weak binding in these cases).

In summary, we have analysed 
energy shifts in pairs of isobaric analogue states (IAS) in mirror nuclei. We find that differences in shifts between pairs of IAS of greater than 27~keV are common, occurring in approximately 35~\% of cases. The inversion of ground- and first-excited-state observed in the recent study~\cite{ref:Hoff_20} of \textsuperscript{73}Sr-\textsuperscript{73}Br lies well within the bounds of ``normal'' Coulombic isospin-symmetry-breaking behaviour and, in the absence of reliable model-predictions contradicting the inversion, requires no additional, exotic symmetry-breaking effects. Finally, we emphasise that state inversions (ground-state or otherwise) provide no more information about isospin-symmetry breaking than any other mirror-energy shift, of which many hundreds have been observed.\\

\noindent{\bf Acknowledgements:} Discussions with P.~Adsley, G.~F.~Bertsch, A.~Gade, R.~V.~F.~Janssens, B.~P.~Kay, T.~D.~Morris, and A.~Ratkiewicz are gratefully acknowledged. Work at LLNL was performed under contract DE-AC52-07NA27344. SRS is supported by the DOE under contract DE-FG02-97ER41014.




\begin{thebibliography}{99}
\bibitem{ref:Bentley_07} M.~A.~Bentley and S.~M.~Lenzi, Coulomb energy differences between high-spin states in isobaric multiplets, Progress in Particle and Nuclear Physics {\bf 59} 497 (2007)
\bibitem{ref:Hoff_20} D.~E.~M.~Hoff, A.~M.~Rogers, S.~M.~Wang, P.~C.~Pender, K.~Brandenburg, K.~Childers, J.~A.~Clark, A.~C.~Dombos, E.~R.~Doucet, S.~Jin, R.~Lewis, S.~N.~Liddick, C.~J.~Lister, Z.~Meisel, C.~Morse, W.~Nazarewicz, H.~Schatz, K.~Schmidt, D.~Soltesz, S.~K.~Subedi and S.~Waniganeththi, Mirror-symmetry violation in bound nuclear ground states, Nature {\bf 580} 52 (2020)
\bibitem{ref:Spieker_19} M.~Spieker, A.~Gade, D.~Weisshaar, B.~A.~Brown, J.~A.~Tostevin, B.~Longfellow, P.~Adrich, D.~Bazin, M.~A.~Bentley, J.~R.~Brown, C.~M.~Campbell, C.~Aa.,~Diget, B.~Elman, T.~Glasmacher, M.~Hill, B.~Pritychenko, A.~Ratkiewicz and D.~Rhodes, One-proton and one-neutron knockout reactions from N=Z=28 \textsuperscript{56}Ni to the A=55 mirror pair \textsuperscript{55}Co and \textsuperscript{55}Ni, Physical Review C {\bf 99} 051304(R) (2019)
\bibitem{ref:ENSDF} Evaluated Nuclear Structure Data File (ENSDF)
\bibitem{ref:Ong_17} W.-J.~Ong, C.~Langer, F.~Montes, A.~Aprahamian, D.~W.~Bardayan, D.~Bazin, B.~A.~Brown, J.~Browne, H.~Crawford, R.~Cyburt, E.~B.~Deleeuw, C.~Domingo-Pardo, A.~Gade, S.~George, P.~Hosmer, L.~Keek, A.~Kontos, I.-Y.~Lee, A.~Lemasson, E.~Lunderberg, Y.~Maeda, M.~Matos, Z.~Meisel, S.~Noji, F.~M.~Nunes, A~.Nystrom, G.~Perdikakis, J.~Pereira, S.~J.~Quinn, F.~Recchia, H.~Schatz, M.~Scott, K.~Siegl, A.~Simon, M.~Smith, A.~Spyrou, J.~Stevens, S.~R.~Stroberg, D.~Weisshaar, J.~Wheeler, K.~Wimmer and R.~G.~T.~Zegers, Low-lying level structure of \textsuperscript{56}Cu and it's implications for the {\it rp} process, Physical Review C {\bf 95} 055806 (2017)
\bibitem{ref:Milne_16} S.~A.~Milne, M.~A.~Bentley, E.~C.~Simpson, T.~Baugher, D.~Bazin, J.~S.~Berryman, A.~M.~Bruce, P.~J.~Davies, C.~Aa.~Diget, A.~Gade, T.~W.~Henry, H.~Iwasaki, A.~Lemasson, S.~M.~Lenzi, S.~McDaniel, D.~R.~Napoli, A.~J.~Nichols, A.~Ratkiewicz, L.~Scruton, S.~R.~Stroberg, J.~A.~Tostevin, D.~Weisshaar, K.~Wimmer and R.~Winkler, Physical Review Letters {\bf 117} 082502 (2016)
\bibitem{ref:Davies_13} P.~J.~Davies, M.~A.~Bentley, T.~W.~Henry, E.~C.~Simpson, A.~Gade, S.~M.~Lenzi, T.~Baugher, D.~Bazin, J.~S.~Berryman, A.~M.~Bruce, C.~Aa.~Diget, H.~Iwasaki, A.~Lemasson, S.~McDaniel, D.~R.~Napoli, A.~Ratkiewicz, L.~Scruton, A.~Shore, S.~R.~Stroberg, J.~A.~Tostevin, D.~Weisshaar, K.~Wimmer and R.~Winkler, Mirror Energy Differences at Large Isospin Studied through Direct Two-Nucleon Knockout, Physical Review Letters {\bf 111} 072501 (2013)
\bibitem{ref:NushellX} B.~A.~Brown and W.~D.~M.~Rae, The Shell-Model Code NuShellX@MSU, Nuclear Data Sheets {\bf 120} 115 (2014)
\bibitem{ref:USDB} B.~A.~Brown and W.~A.~Richter, New ``USD'' Hamiltonians for the {\it sd} shell, Physical Review C, {\bf 74} 034315 (2006)
\bibitem{ref:Honma_05} M.~Honma, T.~Otsuka, B.~A.~Brown, T.~Mizusaki, Shell-model description of neutron-rich pf-shell nuclei with a new effective interaction GXPF1, The European Physical Journal A {\bf 25} 499 (2005)
\bibitem{ref:Ormand_89} W.~E.~Ormand and B.~A.~Brown, Empirical isospin-nonconserving hamiltonians for shell-model calculations, Nuclear Physics A {\bf 491} 1 (1989)
\bibitem{ref:Blomqvist_68} J.~Blomqvist and A.~Molinari, Collective $0^-$ vibrations in even spherical nuclei with tensor forces, Nuclear Physics A {\bf 106} 545 (1968)
\bibitem{ref:Suhonen_07} J.~Suhonen, From Nucleons to Nucleus, Springer-Verlag Berlin Heidelberg, (2007)
\bibitem{ref:Thomas_52} R.~G.~Thomas, An Analysis of the Energy Levels of the Mirror Nuclei, ${\mathrm{C}}^{13}$ and ${\mathrm{N}}^{13}$, Physical Review {\bf 88} 1109 (1952)
\bibitem{ref:Ehrman_51} J.~B.~Ehrman, On the Displacement of Corresponding Energy Levels of ${\mathrm{C}}^{13}$ and ${\mathrm{N}}^{13}$, Physical Review {\bf 81} 412 (1951)
\bibitem{ref:Stefan_14} I.~Stefan, F.~de~Oliveira~Santos, O.~Sorlin, T.~Davinson, M.~Lewitowicz, G.~Dumitry, J.~C.~Ang\'elique, M.~Ang\'elique, E.~Berthoumieux, C.~Borcea, R.~Borcea, A.~Buta, J.~M.~Daugas, F.~de~Grancey, M.~Fadil, S.~Gr\'evy, J.~Kiener, A.~Lefebvre-Schuhl, M.~Lenhardt, J.~Mrazek, F.~Negoita, D.~Pantelica, M.~G.~Pellegriti, L.~Perrot, M.~Ploszajczak, O.~Roig, M.~G.~Saint~Laurent, I.~Ray, M.~Stanoiu, C.~Stodel, V.~Tatischeff and J.~C.~Thomas, Probing nuclear forces beyond the drip-line using the mirror nuclei $^{16}\mathrm{N}$ and $^{16}\mathrm{F}$, Physical Review C {\bf 90} 014307 (2014)
\end{thebibliography}

\end{document}